\documentclass[iop]{emulateapj}
\usepackage{txfonts}
\usepackage{natbib}
\usepackage{subfigure}
\usepackage{url}
\usepackage{graphicx}
\shorttitle{On AGN Outflows Absorption Line Profiles}
\shortauthors{Giustini \& Proga}

\begin{document}

\title{On the Diversity and Complexity of Absorption Line Profiles Produced by Outflows in Active Galactic Nuclei}


\author{Margherita Giustini\altaffilmark{1,2,3} and Daniel Proga\altaffilmark{2,4}}
%



\altaffiltext{1}{INAF-Istituto di Astrofisica Spaziale e Fisica cosmica di Bologna, via Gobetti 101, I-40129, Bologna, Italy.}
\altaffiltext{2}{Department of Physics and Astronomy, University of Nevada Las Vegas, 4505 Maryland Pkwy, Las Vegas, NV 891541-4002, USA.}
\altaffiltext{3}{ Center for Space Science and Technology, University of Maryland, Baltimore County, 1000 Hilltop Circle, Baltimore, MD 21250, USA.}
\altaffiltext{4}{Princeton University Observatory, Peyton Hall, Princeton, NJ 08540, USA.}


\begin{abstract}

Understanding the origin of AGN absorption line profiles and their diversity could help to explain the physical structure of the accretion flow, and also to assess the impact of accretion on the evolution of the AGN host galaxies. Here we present our first attempt to systematically address the issue of the origin of the complexities observed in absorption profiles. Using a simple method, we compute absorption line profiles against a continuum point source for several simulations of accretion disk winds. We investigate the geometrical, ionization, and dynamical effects on the absorption line shapes. We find that significant complexity and diversity of the absorption line profile shapes can be produced by the non-monotonic distribution of the wind velocity, density, and ionization state. Non-monotonic distributions of such quantities are present even in steady-state, smooth disk winds, and naturally lead
to the formation of multiple and detached absorption troughs. These results demonstrate that the part of a wind where an absorption line is formed is not representative of the entire wind. Thus, the information contained in the absorption line is incomplete if not even insufficient to well estimate gross properties of the wind such as the total mass and energy fluxes. In addition, the highly dynamical nature of certain portions of disk winds can have important effects on the estimates of the wind properties. For example, the mass outflow rates can be off up to two orders of magnitude with respect to estimates based on a spherically symmetric, homogeneous, constant velocity wind.

\end{abstract}



\section{INTRODUCTION}

Accretion disk winds are among the most promising physical mechanisms able to
link the small and the large scale phenomena in active galactic nuclei (AGN) and
to shed light on the physics of mass accretion/ejection around supermassive
black holes (SMBHs). Such winds are currently directly observed, as blueshifted
and broadened absorption lines in the UV and X-ray spectra of a substantial
fraction of AGN \citep[e.g.,][]{2003ARA&A..41..117C}. 

In the UV band we observe, with decreasing width of the absorption troughs, broad absorption lines (BALs) in about 15~\% of optically selected AGN \citep[e.g.,][]{2008MNRAS.386.1426K,2009ApJ...692..758G}; mini-broad absorption lines (mini-BALs) and narrow absorption lines (NALs) in about 30~\% of AGN \citep{2008ApJ...672..102G}.
Once corrected for selection effects, the intrinsic fraction of AGN hosting BALs is estimated to be $\sim 30-40$~\% \citep{2011MNRAS.410..860A}; summing also the corrected fraction for mini-BALs and NALs, the intrinsic fraction of AGN hosting blueshifted UV absorption lines results to be as high as $\sim 70$~\% \citep{2012arXiv1204.3791H}. 
In all these cases, the absorbers are associated to resonant transitions of moderately ionized elements like Mg~\textsc{ii}, Al~\textsc{iii}, C~\textsc{iv}, Si~\textsc{v}, and can be outflowing with velocities as high as $\sim 0.2$~$c$, the main difference among them being in the width of their absorption troughs.
Given their large width, BAL structures are the easiest to be identified even in low-resolution UV spectra, and have therefore been studied in much more detail than mini-BAL and NAL structures. 
The shape of BAL troughs is quite complex, showing a variety of profiles among different sources \citep[e.g.,][]{1984ApJ...280...51T, 1987ApJ...317..450F, 1988ApJ...325..651T,2002ApJS..141..267H}. 
When high spectral resolution, high signal-to-noise observations have become available, generally also mini-BALs and NALs showed a variety of profiles \citep[e.g.,][]{2007ApJ...660..152M,2010ApJ...722..997W,2011MNRAS.410.1957H, 2011MNRAS.411..247R}.
Recently a number of AGN have been shown to display strong variability in their UV absorption troughs, including emergence of BAL structures and transitions from mini-BALs to BALs \citep[e.g.,][]{2002MNRAS.335L..99M,2008MNRAS.391L..39H,2009ApJ...701..176L,2010ApJ...724L.203K,2012MNRAS.421L.107V}.

In the X-ray spectra of $\sim 50$~\% of AGN, we observe also lower-velocity (from $\sim 100$ to $\sim 1000$~km~s$^{-1}$) outflowing ionized gas (the so-called warm absorber) in the transitions of ionized elements such as C~\textsc{vi}, O~\textsc{vii}, O~\textsc{viii}, Si~\textsc{vi} \citep[e.g.,][]{2007MNRAS.379.1359M}. 
Higher velocity absorbing gas in the transitions of Fe~\textsc{xxv} and Fe~\textsc{xxvi}, with blueshift in the range $0.003-0.3$~$c$ is observed in a number of AGN \citep[e.g.,][]{2003MNRAS.345..705P, 2005ApJ...630L.129R, 2006AN....327.1012C, 2007MNRAS.375..227M, 2008A&A...483..161T,2011A&A...536A..49G}; their fraction among the AGN population is currently unknown, the only reliable estimate being of about $30-40$~\%
for a sample of low redshift AGN \citep{2010A&A...521A..57T}. 
X-ray BALs outflowing up to $0.7$~$c$ have been also observed in a handful of AGN \citep{2002ApJ...579..169C,2003ApJ...595...85C,2009ApJ...706..644C, 2012A&A...544A...2L}. 
The ionization state of the X-ray absorbers spans a large range of values, from e.g. C~\textsc{vi} in the case of warm absorbers up to Fe~\textsc{xxvi} for X-ray BALs. 
The observed high-velocity X-ray absorbers profiles are quite complex and display strong temporal variability both in depth and velocity shift; also for the case of the lower-velocity X-ray warm absorbers, deep observations have revealed significant complexities in the absorption trough profiles \citep[e.g.,][]{2005ApJ...627..166R,2011MNRAS.413.1251P} and variability in ionization and velocity shift \citep[e.g., see][for some recent results]{2010A&A...510A..92L,2011A&A...533A...1M}.

Velocity and ionization state are only two of the physical properties of AGN
winds; the mass and energy fluxes carried by them are other fundamental
properties that we wish to be able to measure in order to quantify the actual
impact of these winds on the surrounding environment, i.e., the amount of
feedback. To measure the mass and energy flux, the density of the wind
material and the distance between the central SMBH and the place where
absorption occurs must be known. Unfortunately, these two quantities are
difficult to measure, and estimates for them have always relied on several
assumptions. In particular, a geometrically and optically thin,
constant-density, spherically symmetric, single-zone of gas in ionization
equilibrium and outflowing with a uniform velocity has been usually assumed to derive
constraints on the distance of the absorbing material from the central ionizing
source.

Using these assumptions, reported estimates of the distances of absorbers from
the central SMBH range from the inner regions of the accretion disk for the
high-velocity X-ray BALs \citep[e.g.,][]{2009ApJ...706..644C}, to the parsec-scale torus for
the X-ray warm absorbers \citep[e.g.,][]{2005A&A...431..111B}, to the kiloparsec-scale for some
UV BALs \citep[e.g.,][]{2010ApJ...709..611D}. Currently the uncertainties on the distances are
relatively large, which translates to a high uncertainty on the mass outflow
rate and on the kinetic energy injection associated with such winds \citep[e.g.,][]{2003ApJ...593L..65R,2007ApJ...659.1022K,2009A&A...504..401C}.

Theoretical arguments and numerical simulations have been used to show that
accretion disks are able to launch and accelerate powerful outflows by various
physical mechanisms such as radiation pressure, magnetic pressure, and thermal
pressure \citep[see e.g.][for reviews on the subject]{2006MmSAI..77..598K,2007ASPC..373..267P, 2007Ap&SS.311..269E}. The driving
mechanism and the accreting system physical parameters (e.g., accretion rate,
black hole mass, magnetic field configuration, UV/X-ray flux ratio) determine
the wind characteristics \citep[e.g.,][]{2005ApJ...630L...9P}. Despite the many parameters that
combine themselves in complicated ways, only a few geometries of streamlines are
shown by AGN accretion disk wind models.

\begin{figure} 
\figurenum{1} \epsscale{.9} \plotone{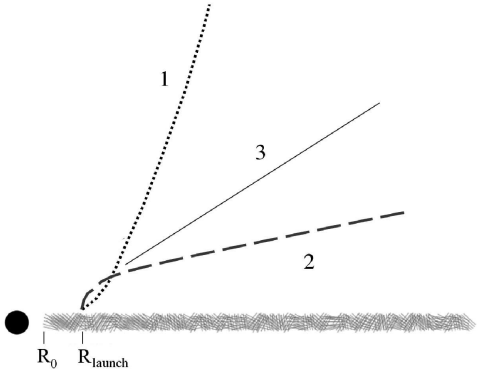} 
 \caption{\label{f1}The three ``typical'' streamlines
for accretion disk winds. 1: MHD-typical, polar and convex streamline.
2: hydrodynamical-typical, equatorial and concave streamline. 3: a radial
streamline typical of spherically symmetric winds.} 
 \end{figure}
 
Figure~\ref{f1} shows the main general shapes of streamlines that are predicted
by models of AGN accretion disk winds. Number 1 is a convex, polar streamline,
typical of magnetic driving scenarios \citep[e.g.,][]{1982MNRAS.199..883B,1991ApJ...379..696L,2003ApJ...595..631K, 2005ApJ...630..945A, 2005ApJ...631..689E}. Number 2 is a
concave, more equatorial streamline, typical of line-driven
\citep[e.g.,][]{2000ApJ...543..686P, 2004ApJ...616..688P} and thermally driven disk winds \citep[e.g.,][]{1996ApJ...461..767W,2004ApJ...607..890F, 2010ApJ...719..515L}. What is usually assumed in order
to deduce quantities such as mass outflow rate, ionization state, or absorber
distance from the continuum source is the configuration number 3, i.e. a radial
flow. 

However, the different geometries correspond to very different wind
properties: the mass flux $\dot{m}_{out}$ across the area of a flow tube, $A(r)$
is 
\begin{equation}\dot{m}_{out}= A(r) \upsilon(r) \rho(r)\end{equation} 
where $\upsilon$ is the velocity
normal to the area, and $\rho$ is the density. For a simple radial wind with a
constant velocity, $A(r) \propto r^2$ and $\rho(r)\propto r^{-2}$, whereas for a
cylindrical wind (with constant velocity) $A(r)=const$ and $\rho(r)=const$. This
density dependence on the flow geometry has received little consideration in
previous works on estimating the flow mass and energy rates using observations
\citep[however, see][]{2012arXiv1207.7348W}.

In this article we will report on results from our first attempt of
systematically addressing the following questions: what drives the appearance of
different line profiles? Can we distinguish among different geometries for
accretion disk winds from observations? How reliable are estimates of wind
properties, when a radial or spherical flow is assumed? The secondary questions
that will lead to answer the former ones are then: what are the implications of
radial flow assumptions when deriving physical quantities related to the wind?
Do different wind physical characteristics produce different line profiles? If
yes: how and why? Our method to compute line profiles is introduced in
Section 2; results are presented in Section 3 and discussed in Section 4; finally, conclusions are presented in
Section 5.

\section{METHOD}

We consider several examples of disk wind models, with an increasing level 
of completeness or complexity or both. We present results in terms of 
synthetic absorption line profiles predicted by these models.
Specifically, we first consider a simplified, isothermal, thermally driven 
disk wind model and use steady state solutions as predicted by hydrodynamical 
simulations similar to those of
\citet{2004ApJ...607..890F} in Section~\ref{geometry}; 
then we consider hydrodynamical simulations as given by 
\citet{2010ApJ...719..515L}, where the effect of a central X-ray illuminating
source is explicitly taken into account, showing the effects of 
varying ionization state across the flow (Section \ref{luketic}); 
finally we consider the solutions given by \citet{2004ApJ...616..688P}, 
showing the effects of the departure of the flow
from a steady configuration (Section \ref{proga}).

Our method to compute absorption line profiles adopts a radiation transfer 
treatment that is simplified to the bare minimum, nonetheless capturing 
the most important physical and geometrical effects, and as we will show 
is useful to explore the effects of different wind geometries. Specifically, 
we ignore scattered radiation and we assume that the continuum photons 
are emitted by a point source. Therefore we deal with proper 
``line of sight'' (LOS), instead of ``cylinder of sight'' and
line profiles observed at infinity are sensitive only to
the radial component of the wind velocity only along a given LOS.
This kind of approach is viable for analyzing e.g. X-ray properties of AGN 
winds because of the compactness of the X-ray source continuum. 
Having the distributions of the density $\rho$ and the radial velocity 
$\upsilon_r$ for each wind model, we compute synthetic absorption line 
profiles corresponding to different LOS or equivalently inclination angles. 
Our procedure is the following: First, we search for resonance points 
given the continuum frequency, $\nu_o$ along 
a given LOS using the resonance condition
\begin{equation} 
\frac{\left(\nu - \nu_o\right)}{\nu_o} =\frac{\upsilon_r}{c}
\end{equation}
where $\nu$ is the Doppler shifted flow frequency with respect to the continuum, 
and $\upsilon_r$ is the velocity along the LOS. 
We then use the Sobolev approximation to compute the optical depth:
 \begin{equation}\label{eq3}
 \tau = \frac{\kappa c}{\nu}\rho\frac{1}{|d\upsilon_r(r)/dr|}
 \end{equation} 
where the line opacity $\kappa$ is a product of the oscillator
strength and the abundance of a given element and ionization factor for a given ion,
and $d\upsilon_r(r)/dr$ is the velocity gradient along the LOS
(with our approximation it is just the gradient of the radial velocity).
To focus on the effects of the density and radial velocity distributions across the flow,
we set $\kappa c/\nu=1$ and compute the optical depth $\tau_{\nu,i}$ at a given resonance point $i$, 
and the total optical depth, $\tau_{\nu}$  at a given frequency
by summing up the optical depths at all resonance points that we found
\begin{equation}
\tau_{\nu}=\sum_{i}\tau_{\nu,i}
\end{equation} 
We finally solve the radiative transfer equation:
\begin{equation}
I=I_o e^{-\tau_{\nu}}\rm{.} 
\end{equation}

Many already studied how synthetic line profiles depend on the wind geometry 
and viewing angle. However, most of previous studies used {\it kinematic} 
models of accretion disk winds and were applied to cataclysmic variables 
\citep[e.g.,][]{1987MNRAS.224..595D, 
1993ApJ...409..372S,1995MNRAS.273..225K,2002ApJ...579..725L}. These models
were axisymmetric. They assumed streamlines to be {\it straight} lines
and the wind velocity to be some prescribed function of the position along 
the streamline. 
By varying one of the free parameters
(namely the inclination angles that the streamlines made with respect to the disk plane),
such models permit the exploration of various wind geometries, 
but only to some extent. 
One of the results of these studies is that the formation of
blueshifted broad absorption troughs (i.e., P Cygni profiles) can be naturally
obtained for a variety of wind geometries. 
Our work is an extension of those earlier studies as we 
investigate how the absorption troughs depend on the geometry of 
the streamlines, the opacity, and the ionization state but instead of 
using kinematic models we use solutions of hydrodynamical equations
obtained by numerical simulations. More sophisticated calculations
of wind photoionization and radiative transfer are possible, however they
are relatively time consuming and therefore less suitable
to the exploration of various geometries 
\citep[e.g.,][]{2010MNRAS.408.1396S}.

\section{RESULTS}\label{RES}

\begin{figure*} 
\begin{center}
\figurenum{2} \includegraphics[height=5.5cm]{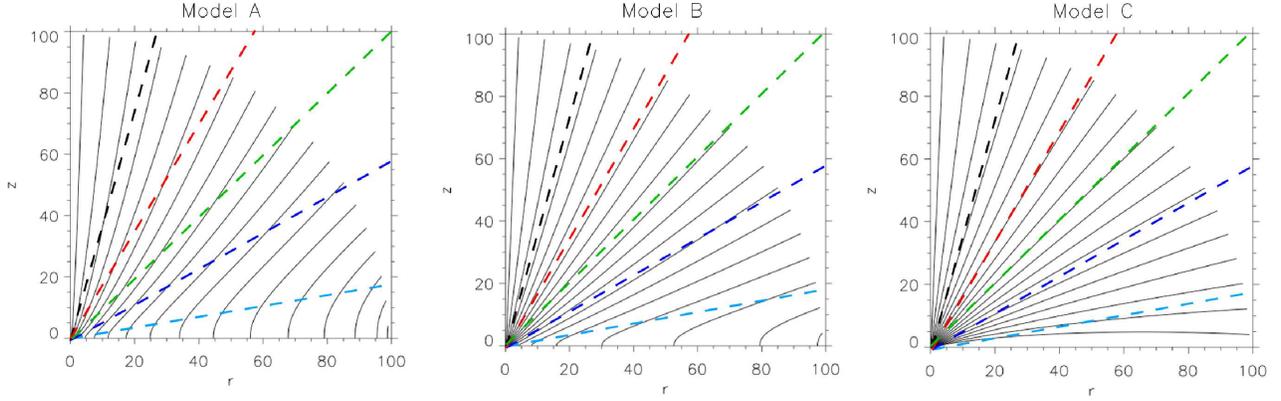}
 \caption{\label{f2}Streamlines for the three simple isothermal disk wind models 
adopted from \citet{2010ApJ...719..515L} \citep[see also][]{2004ApJ...607..890F}.
The models were computed  assuming a density profile at
the wind base as the following function of radius, $r_D$: 
$\rho(r)\propto r_D^{-\alpha}$ for $\alpha=1.5, 2.0,$ and  2.5, from left to right. 
The colored dashed lines represent the five LOS for which we calculate absorption
line profiles: $i =15, 30, 45, 60,$ and 80$^{\circ}$ for the black, red, green, dark blue, and light blue line, respectively. 
The length scales on the figure are in units of the radius at which the gas internal energy equals
the gravitational energy of the central object located at (0,0).} 
\end{center}
 \end{figure*}

\begin{figure*} 
\figurenum{3} 
\subfigure{\includegraphics[width=5.7cm]{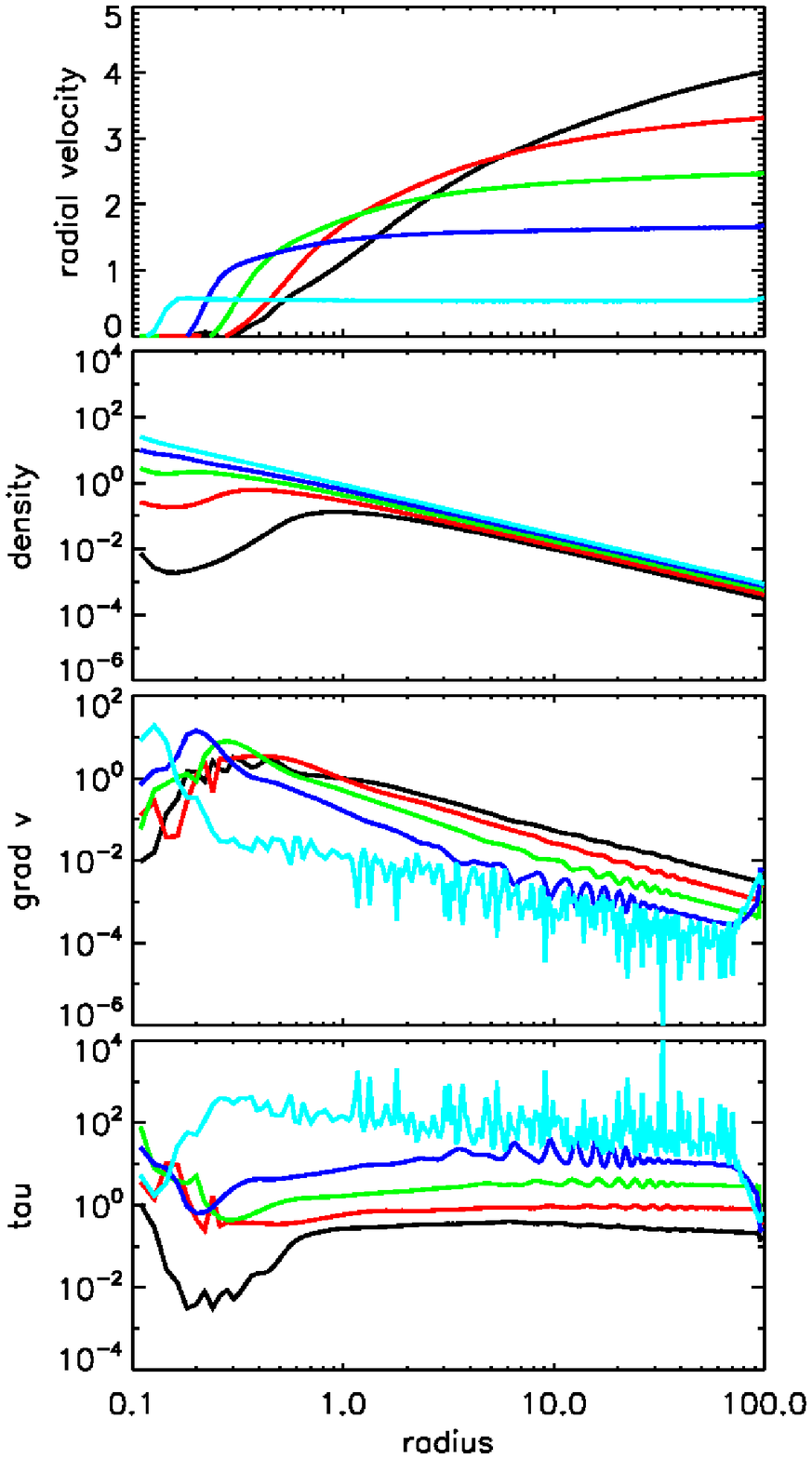}\includegraphics[width=5.7cm]{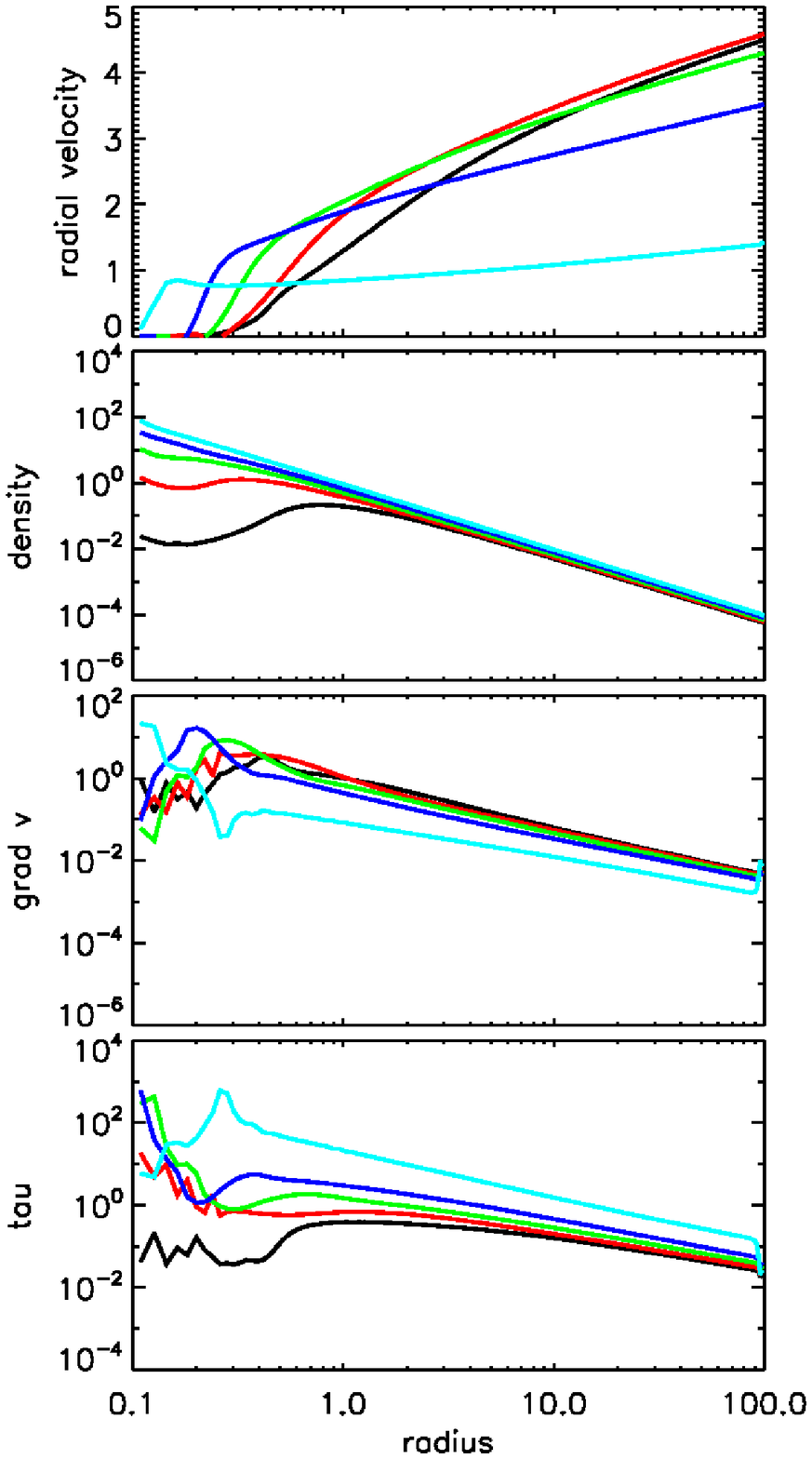}\includegraphics[width=5.7cm]{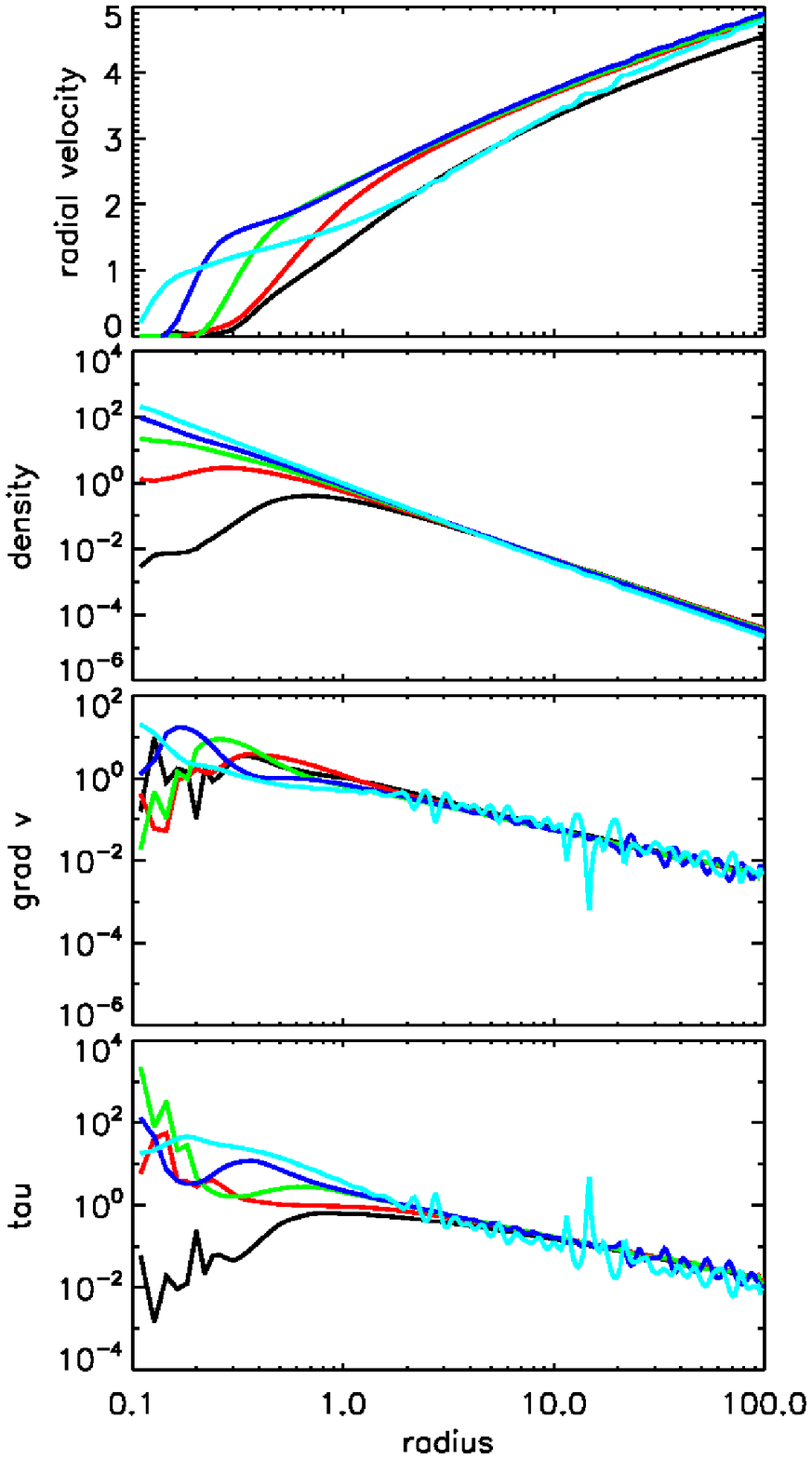}}
\caption{Radial profiles for the velocity, density, radial velocity gradient, and finally line optical depth, for model A, B, and C from left to right. The color code is as in Figure~\ref{f2}. Radii are in units of the radius at which the gas thermal energy equals the gravitational energy of the central object, while velocities are in units of sound speed.\label{f3}}
\end{figure*}
\begin{figure*} 
\figurenum{4} \subfigure{\includegraphics[width=5.7cm]{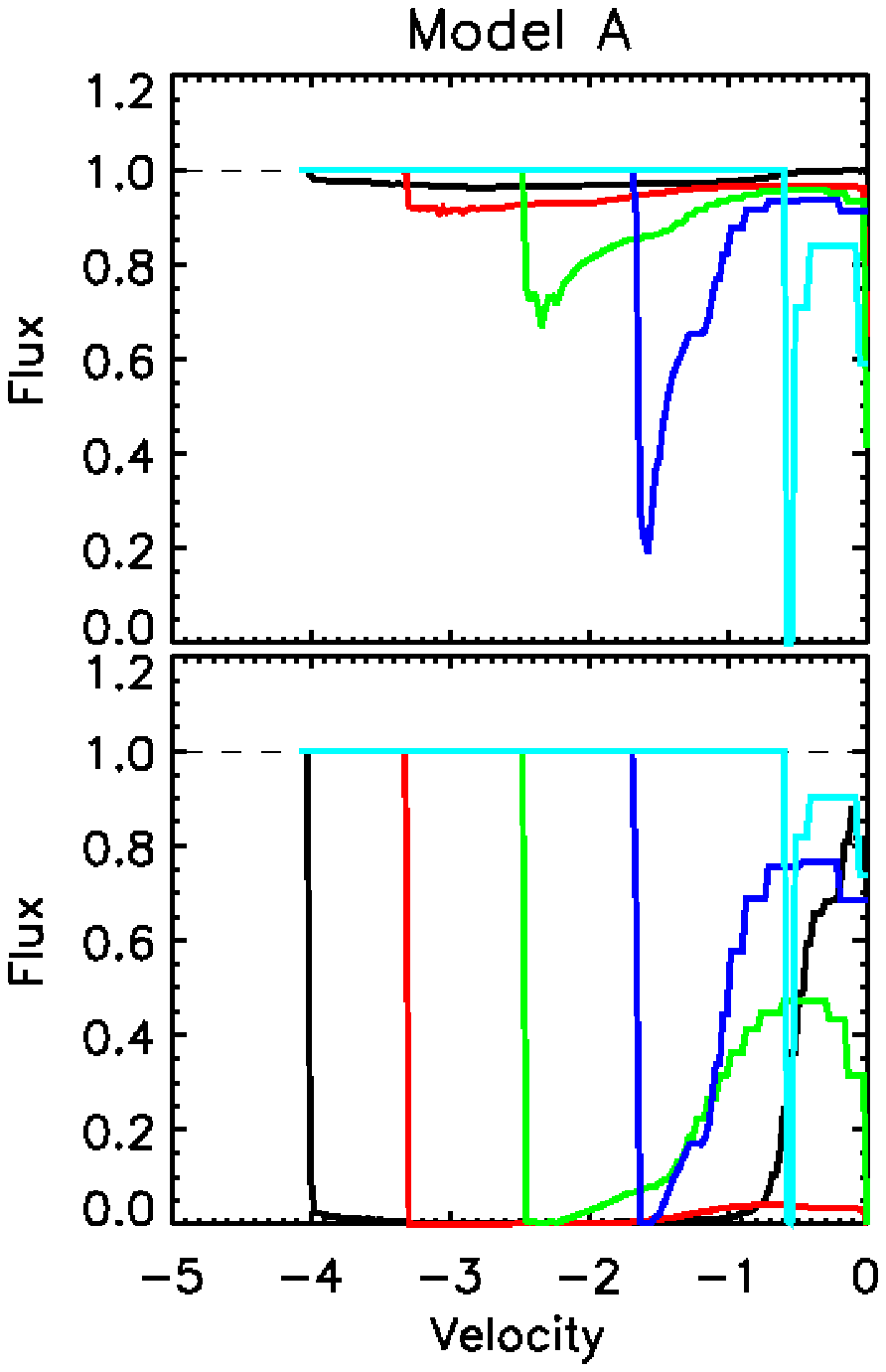}\includegraphics[width=5.7cm]{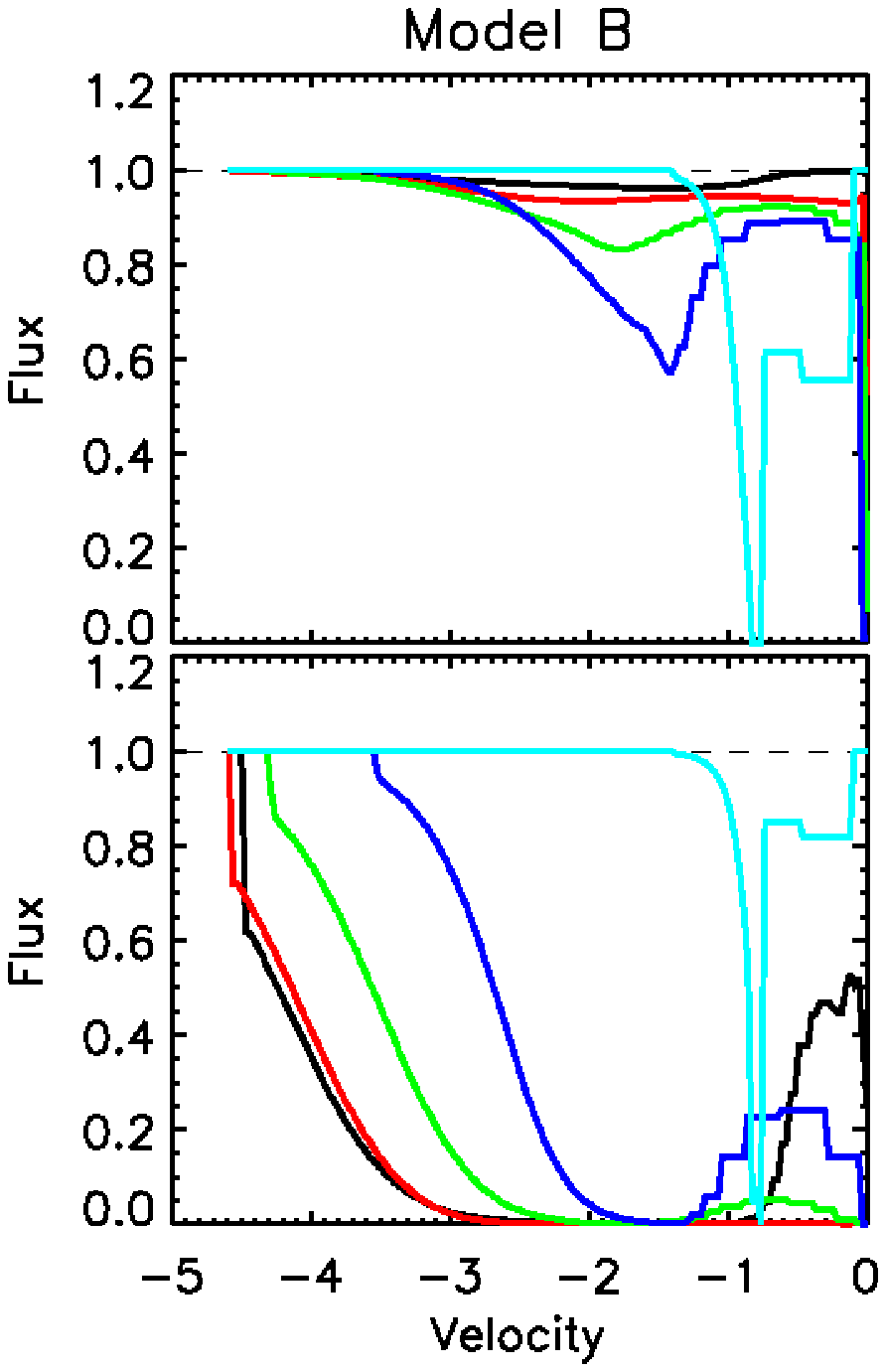}\includegraphics[width=5.7cm]{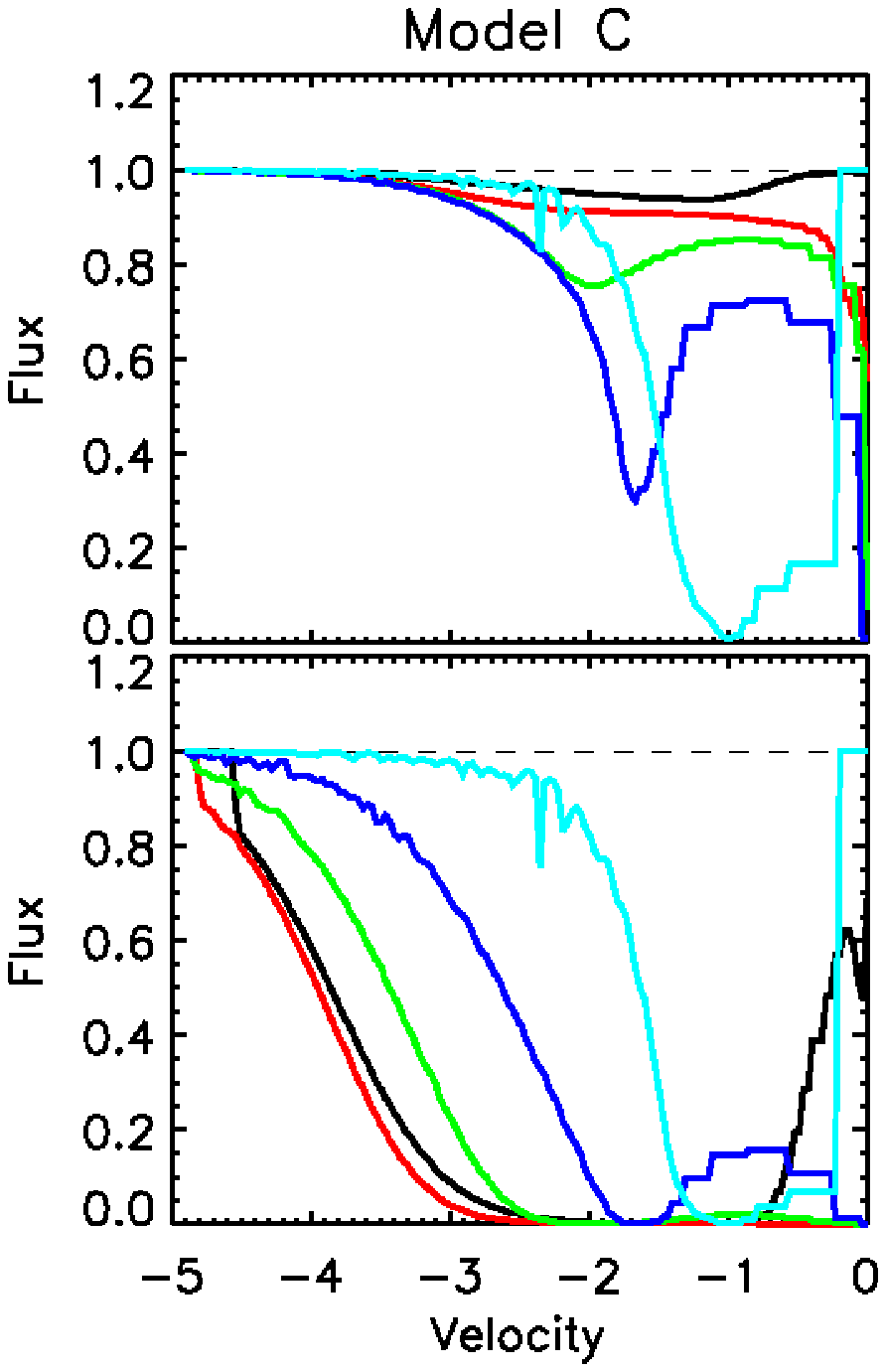}}
\caption{Absorption line profiles for the for each model, along the five LOS color-coded as in Figure~\ref{f2}.
Top panels: profiles computed using the actual opacity of the models; bottom panels:
profiles computed increasing the opacity for each LOS until the zero flux level
is reached at least in one point in the velocity space.\label{f4}}
\end{figure*}

\subsection{Effects of the Overall Wind Geometry\label{geometry}}

We first consider 2D, axisymmetric, isothermal, steady state disk wind models. 
We use models computed by \citet{2010ApJ...719..515L} and refer a reader 
to this work for details about the simulations. We note 
that \citet{2010ApJ...719..515L} performed these simulations to test
their code and setup against results presented by 
\citet{2004ApJ...607..890F}.

The latter showed that  a relatively wide range of the wind geometries and 
shapes of the streamlines can be modeled by varying the slope of the radial 
profile of the density along the disk atmosphere (the base of the wind). 
In particular, Figure~\ref{f2} shows examples of three different wind geometries
assuming a power-law density profile:  
$\rho(r)\propto r_D^{-\alpha}$ and three values of $\alpha$: 
1.5 (model A), 2.0 (model B), and $2.5$ (model C).
As found by \citet{2004ApJ...607..890F}, all the three models predict a 
very similar shape and divergence (i.e., $A(r)$) for the streamlines
originating from the inner most part of the disk;
for other streamlines (or middle and high inclination angles), 
some differences and trends among various models start to appear. 
The flattest radial density profile along the wind base
(model A) corresponds to a relatively polar geometry,
while the steepest profile (model C) is the one that most resembles 
the spherical outflow case (quasi radial, with spherical divergence). 

We probed the winds along five different LOS, indicated with the colored dashed
lines in Fig.~\ref{f2}; the radial velocity, density, radial velocity gradient,
and finally optical depth are plotted in Fig.~\ref{f3} for each model and for
each inclination angle. All the quantities are arbitrarily normalized, as at
this point, we simply want to explore the qualitative behavior of the different
geometries: to this end, what is relevant is the relative contribution of the
different parts of the flow and not their absolute magnitude, the latter being
strongly dependent on the actual details of the system such as the central mass,
the Eddington ratio, and the spectral energy distribution.

Figure~\ref{f4} shows the resulting absorption line profiles using the same 
color code as in the previous figures. To show the profile shapes in more detail, the bottom panels 
of Figure~\ref{f4} show results where the opacity has been multiplied by a factor so 
that the point in velocity space where the opacity is maximum reaches the zero flux level for each LOS.

The most obvious difference between the profiles predicted by winds of 
different geometries is in the shape of the blue wing:
very sharp for model A and  very smooth for model C. Model B is an intermediate
case: similar to model A at high inclination angles ($i \sim 80^{\circ}$),
to model C at middle and low inclination angles ($i \lesssim 60^{\circ}$). 
The main effect of $i$ on the profiles, regardless of the wind model, 
is on the profile width: the larger the inclination the narrower the profile. 
However, the profile sensitivity to $i$ is model dependent. In particular,
the profile width is most sensitive to $i$ in model A and the least
in model C. This is simply a reflection of the fact that model~A is polar
whereas model~C is almost spherically symmetric. The right panels of
Figure~\ref{f3} show that for Model~C, at large radii, the flow properties
do not depend on $i$ (profiles for various $i$ collapse
onto one power-law) as expected for a spherical solution, 
whereas the panels in the left and middle columns
show that the flow properties are strong functions of $i$.

For the highest inclination angle (light blue lines in Figure ~\ref{f3}) in Model~A
and C, the radial profiles for the opacity and the radial velocity gradient appear to be
noisy. In the case of Model A, this is a numerical effect which is related to the nearly
constant velocity, that makes its gradient very small; as for Model~C, the noise
reflects the fluctuations in the radial velocity profile close to the base of the wind
(i.e. the disk surface).

In this subsection, we consider the isothermal disk wind solutions. 
One of the key 
consequences of assuming constant temperature is that the wind velocity 
along a given streamline is an ever increasing function of the radius. 
In other words, those models have infinite acceleration zones. This can be 
seen most clearly in the top right panel of Figure~\ref{f3} showing the radial 
velocity for model~C and various $i$. In model~C, the flow is almost
radial so that the velocity plotted in Figure~\ref{f3} (the radial component)
is very similar to the total velocity. Note that this is not the case 
for the plotted velocity of model~A that could give an impression that 
the velocity saturates at a finite level. This saturation
is not caused by a finite acceleration but 
by projecting the total velocity onto a LOS (in effect, taking only 
the radial component of the total velocity). This projecting significantly 
affects the plotted velocity for model~A and to some extent for model~B. 

A more detailed analysis shows that for the case of polar flows (model A), 
the optical depth is a weak function of radius at large radii. This
explains a relatively flat line profiles at large velocities and 
a sharp edge corresponding to the velocity at $r_{out}$. The flatness
of the $\tau$ profile is caused by the relative
flatness of the density and radial velocity profiles.
In the case of more equatorial flows (model B and C), the density 
scales as expected for the spherical flow like $r^{-2}$ and $\tau$
is a decreasing function of radius at large radii. Therefore, the line
absorption reaches a maximum at some intermediate velocity
and then it gradually decreases with increasing velocity with the signature
of the finite size of the computational domain (the sharp
edge) being weak or undetectable. As we will illustrate in next section,
sharp edges or features in the line profiles can be produced
not only as artifacts of the finite computational domain but
also as physical consequences of the internal structure of the wind
\citep[see also the work of][]{1987MNRAS.224..595D, 1993ApJ...409..372S, 1995MNRAS.273..225K}.

Although the profile width depends on $i$, its overall shape is not very 
sensitive to $i$ for these simple isothermal models:
the differences between profiles for different wind geometries seen at
different inclination angles
are not very significant - most  profiles are flat, featureless,  and
differ only in  the position of the blue edge.
In particular, looking at the bottom panels of Figure~\ref{f4}, it is really hard
to distinguish among model B and C, and also among the different
inclination angles. For example, the line profile for $i = 60^{\circ}$
of model B is very similar to the profile for $i = 80^{\circ}$ of
model C. Looking at model A, the profile shapes are now undistinguishable
among different LOS, the only difference being in
the maximum blueshift velocity and the width of the absorption
trough. Given the very similar characteristics of the synthetic absorption
line profiles resulting from different inclination angles
and different geometries, we conclude that different wind geometries
are not the main contributors to the very different shapes of
observed absorption lines.

\subsection{Effects of Finite Acceleration Zone and of Position Dependent Ionization State\label{luketic}}

\begin{figure} 
\begin{center}
\figurenum{5}
 \includegraphics[width=5.5cm]{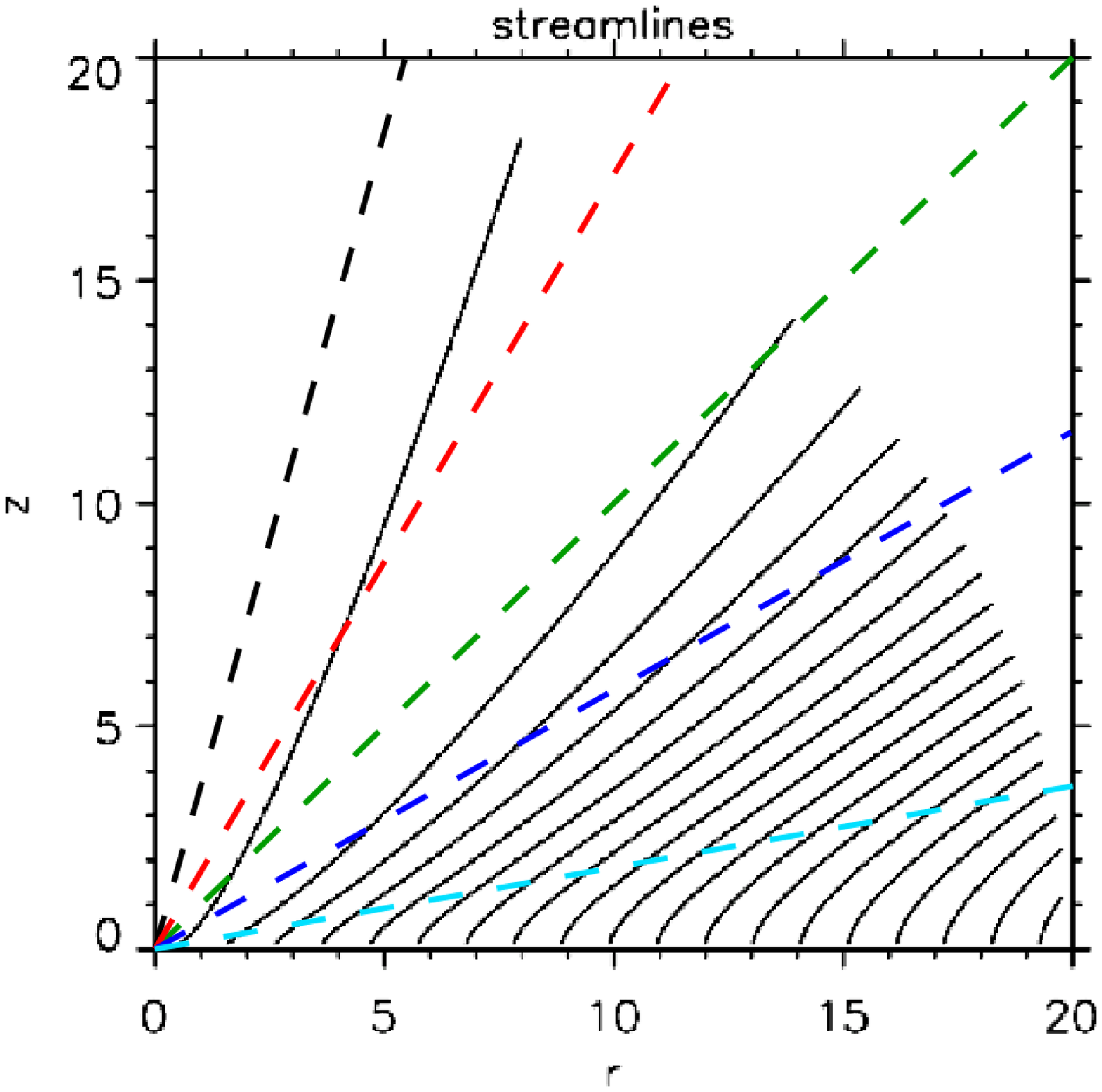}\\
 \includegraphics[width=6.cm]{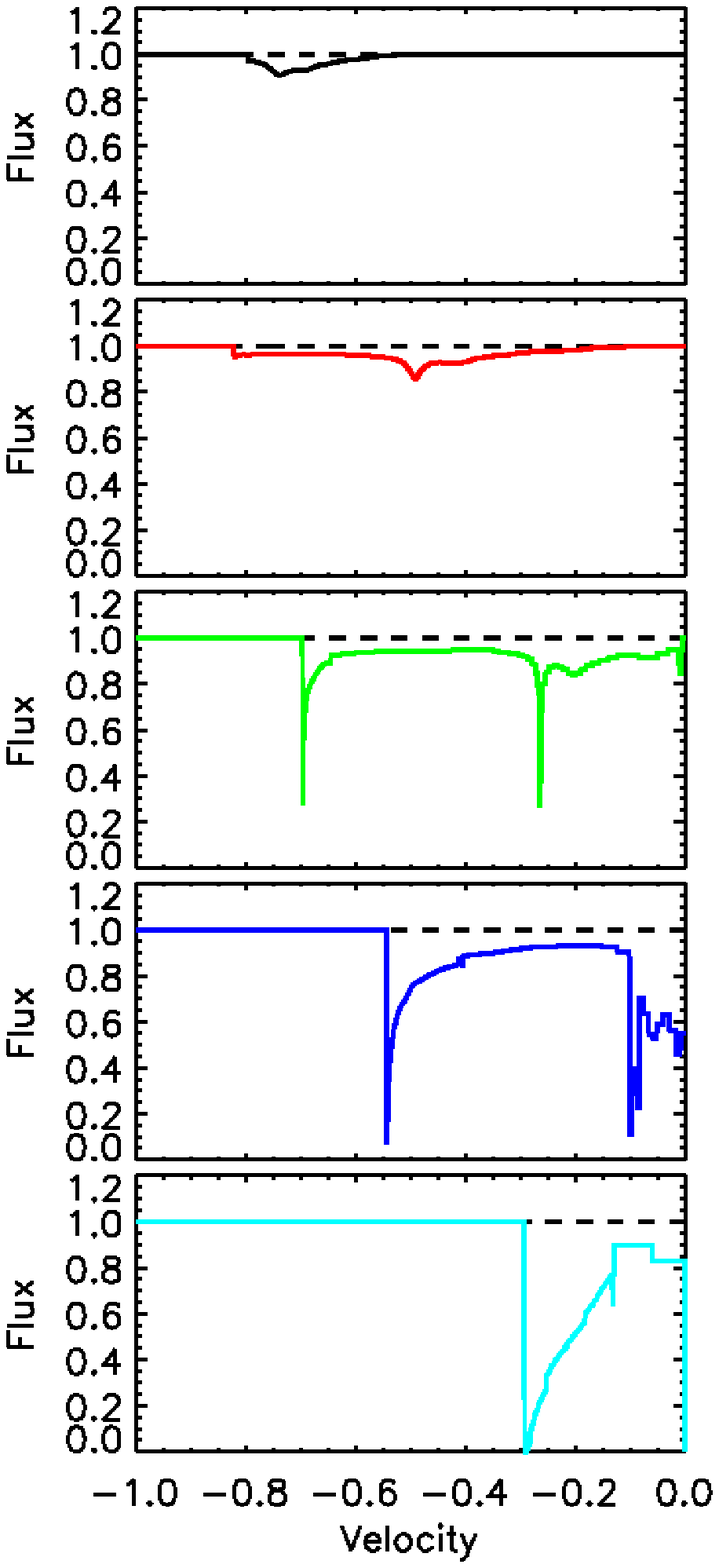}\\
 \caption{\label{f5}Top panel: streamlines for the disk wind model of \citet{2010ApJ...719..515L};
bottom panels: absorption line profiles for different inclination angles, with the same color code as in 
the top panel, for the rerun of the C8 model of \citet{2010ApJ...719..515L}. The velocity scale has been normalized to the maximum 
outflow velocity $\upsilon_{max}=6.6\times 10^7$ cm s$^{-1}$.} 
\end{center}\end{figure}

\begin{figure}\begin{center}\figurenum{6}
 \includegraphics[width=7cm]{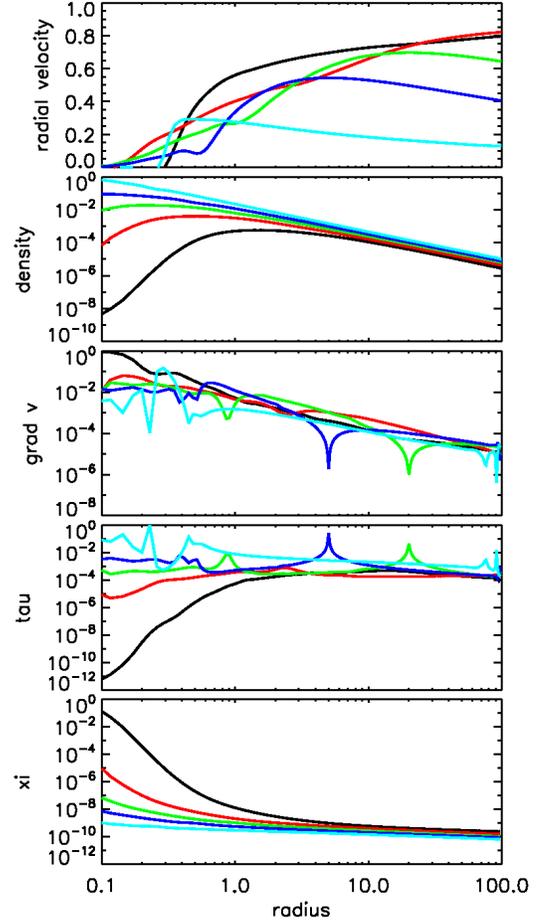}
 \caption{Radial profiles for the velocity, density, radial velocity gradient, line optical depth, and ionization parameter, for the rerun of the \citet{2010ApJ...719..515L} model. The color code is as in 
Figure~\ref{f5}; the quantities have been normalized to their maximum value among all the LOS.\label{f6}} 
\end{center}\end{figure}

\noindent In the previous section, we found that as simple and easy 
to understand the isothermal models are, they are of a limited
usage and could be a source of artifacts in the line profiles. 
For example, the position of the blue edge in the profile for 
model~A would shift to higher velocities if the outer radius of 
the computational domain were increased.
This shift would be a consequence of the fact that
the computation would capture a faster part of the wind solution.
For these and other reasons, we consider next a model for which the assumption
of isothermal flow was relaxed and instead the gas temperature was computed
by introducing physically consistent radiative cooling and heating
processes. We use one of the simulations of the thermally driven
disk wind from \citet{2010ApJ...719..515L}. Specifically, we use
a rerun of their fiducial run (C8) on a grid with an outer radius 
$r=100r_{C}$, where $r_C$ is the Compton radius. 
For most of their runs, \citet{2010ApJ...719..515L} used a grid with a smaller outer radius ($r=20r_{C}$). However,
 to examine and confirm the self-similarity of the flow at large radii, they performed several simulations on the bigger grid.
Figure~\ref{f5} reports the geometry of the wind and the five probed LOS in its top panel, and
the corresponding absorption line profiles in its bottom panels. The difference in profiles 
with respect to the previous case is striking. In this case, the inclination 
angle $i$ plays a fundamental role in shaping the absorption line profiles.
Generally, the profiles are no smooth nor broad anymore, and show two 
distinct components for a range of $i$. 

To associate the line profile characteristics with the wind properties,
Fig.~\ref{f6} shows the radial profiles of the radial velocity, 
the density, the radial velocity gradient, the optical depth, and the ionization parameter
for five $i$'s. The main difference compared to 
the previous model is that the radial velocity reaches a maximum or two 
within the computational domain for most $i$ and
after reaching the maximum,
the radial velocity can significantly decrease with increasing radius 
(see the top panel of Fig.~\ref{f6}). 
This results in large velocity gradients and in turn sharp
maximums in $\tau$.

However, the line forming region can be smaller than the size of the wind,
so that only one of the line components can be produced by a real system.
This can occur in winds where changes in the flow ionization limit the
formation of a given line to a relatively small region.
To illustrate this point, we will scale the ion fraction
with the photoionization parameter
$\xi=L/4 \pi n r^2$ where $L$ is the ionizing luminosity, $n$ is the number
density of the gas, and $r$ is the distance from the continuum source.
Only in  special cases $\xi$ is constant, for example
when $n \propto r^{-2}$ and the flow is radial and of constant velocity. 
For a radial flow with variable velocity, $\xi \propto \upsilon(r)$ so
$\xi$ increases with increase outflow velocity 
\citep[e.g.,][]{2005ApJ...627..166R}.
However, for most $i$, in all the disk wind models we discuss here,
$n$ decreases with $r$ slower than $r^{-2}$ and could
even increase with $r$. Consequently, the photoionization parameter
decreases with increasing radius for a given LOS.

\begin{figure} 
\begin{center}
\figurenum{7}
 \includegraphics[width=7.cm]{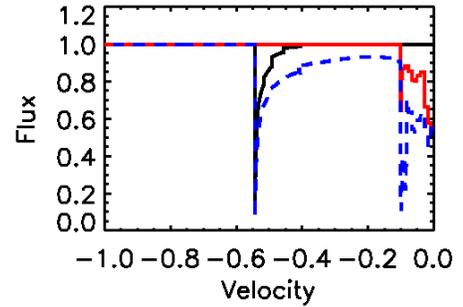}\\
 \caption{\label{fxi}The effect of varying 
ionization parameter for the $i =60^{\circ}$ LOS: in black the lower 
ionization state species, in red the higher ionization state species, in blue 
dashed line the total line profile without accounting for ionization parameter 
distribution effects.} 
\end{center}\end{figure}

In Figure~\ref{fxi}, we show the effects of the varying 
photoionization parameter through the flow, on the absorption line 
profile for $i = 60^{\circ}$. 
Specifically, we recalculated the profile 
assuming that the number density of a given species is a function of $\xi$:
we used two thresholds with a difference in $\xi$ of about two orders of 
magnitude (e.g., to mimic the difference between O~\textsc{viii} and Fe~\textsc{xxv}). 
The profile without effects of $\xi$ is plotted using a blue dashed line 
(it is the same profile as in Figure~\ref{f5} plotted using a blue solid line, 
and it is double-peaked reflecting the two maxima in the opacity distribution along this LOS
visible in Figure~\ref{f6}), while
the profiles with the effects of the ionization state are:
the red line refers to a highly ionized phase and
the black line refers to a lowly ionized one. After taking into account the ionization 
state distribution across the flow, one can easily lose the double-peakness and
find a single absorption trough for each ionic species.
Furthermore, as a consequence of the radial distribution of $\xi$ through the flow, 
one can have a greater blueshift for the lowly ionized species than 
for the highly ionized ones. This is an opposite trend to the one
expected for a radial outflow! 

Under the assumption of a simple radial flow, to explain the presence of 
multiple absorption line profiles that display complexities both in 
the velocity and ionization state parameter spaces, 
one typically is pressed to invoke a multiple-phase absorbing gas
at different locations and with different velocities. 
However, by relaxing the radial 
flow assumption and adopting a more realistic scenario,
these complexities are natural consequences of the geometrical
extent of the outflow and proper treatment of cooling/heating processes.

\subsection{Effects of Unsteady Behavior \label{proga}}

\begin{figure}\figurenum{8}
\begin{center}\includegraphics[width=6cm]{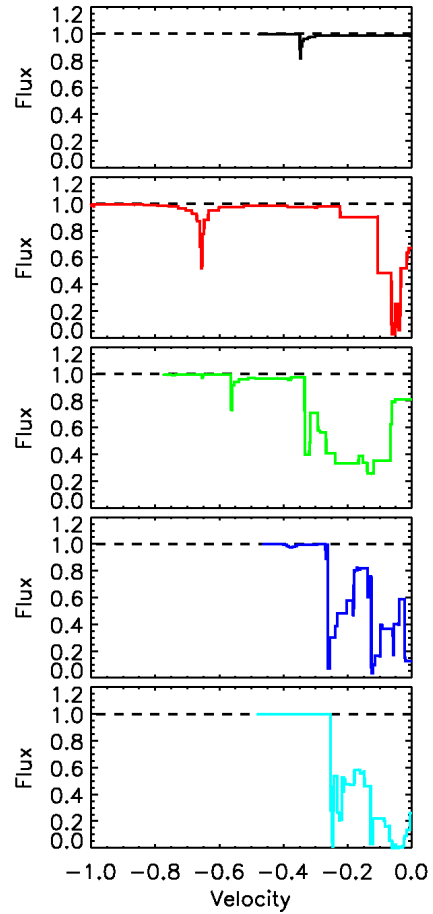}
\caption{ \label{f7}Blueshifted absorption line profiles computed for the LD accretion 
disk wind model of \citet{2004ApJ...616..688P} as seen by the different 
inclination angles $i =$ 60, 65, 70, 75, and 80$^{\circ}$ from the disk 
rotational axis for the black, red, green, blue, and light blue, 
respectively. The velocity scale has been normalized to the maximum 
outflow velocity $\upsilon_{max}=2.9\times 10^9$~cm~s$^{-1}$.}
\end{center}
\end{figure}

\begin{figure}
\begin{center}
\figurenum{9}
\includegraphics[width=7cm]{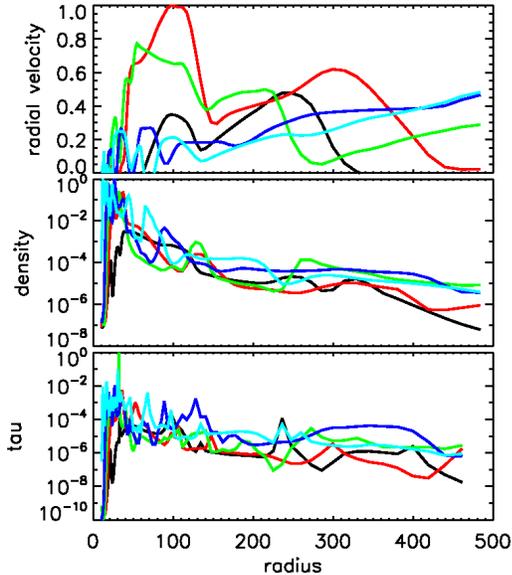}
\caption{\label{f8}From top to bottom velocity, density, and optical depth profiles 
as a function of radius, for the same inclination angles color coded in 
Figure~\ref{f7}. The radius in in units of the inner disk radius, $R_0=3R_S$ where 
$R_S=2GM_{BH}/c^2$ is the Schwarzschild radius \citep[see][for details]{2004ApJ...616..688P}.}
\end{center}
\end{figure}
\begin{figure}
\begin{center}
\figurenum{10}
\includegraphics[width=7cm]{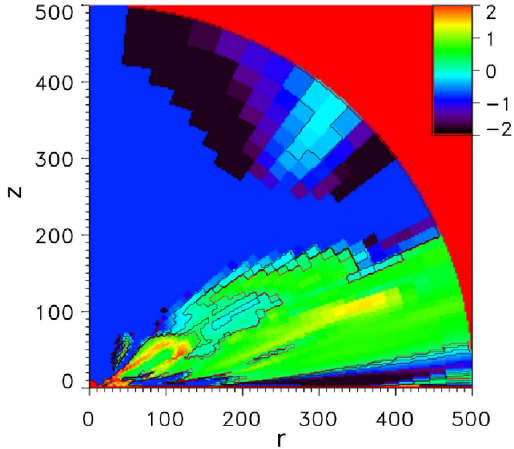}
\caption{\label{f9}Map of the ratio between the mass outflow rate computed 
under the spherical symmetry assumption, and by using actual values from 
the numerical simulations of \citet{2004ApJ...616..688P}. The axes are in units of the inner disk radius, $R_0=3R_S$ where 
$R_S=2GM_{BH}/c^2$ is the Schwarzschild radius.}
\end{center}
\end{figure}

So far, we have considered time-independent solutions of disk wind models. To
illustrate the effects of unsteady solutions, we refer to the line-driven (LD) accretion disk wind
model as presented by \citet{2004ApJ...616..688P}. These authors presented
hydrodynamical simulations of a quasar, demonstrating how the accreting system is
able to launch powerful equatorial disk winds. In particular, they found three
different regimes for the flow: a 'hot flow' in the polar region, where the
density and the velocity are very low; a 'fast stream' in the equatorial region,
where a wind successfully developes, and a 'transitional zone' in between the
two, where the flow is hot and struggles to escape the system. The 'transitional
zone' is effectively shielding the 'fast stream' flow from the strong central
ionizing continuum, preventing it from becoming over-ionized and losing
acceleration in the UV resonant absorption lines. Broad-band X-ray spectra
computed using detailed Monte Carlo simulations to treat the radiative transfer
in the flow are presented in \citet{2010MNRAS.408.1396S}. 
Here we limit our study to the analysis of
simple absorption profiles computed under the Sobolev approximation as in the
previous Sections, using as input the flow physical parameters as given by the
\citet{2004ApJ...616..688P} hydrodynamical simulations.

In Figure~\ref{f7}, we show the absorption line profiles computed for five
different inclination angles through the 'fast stream' and the 'transitional
zone' of the flow, $i =60, 65, 70, 75,$ and $80^{\circ}$ from top to bottom.
Figure~\ref{f8} shows the radial profiles of the velocity, density, and optical
depth for the same inclination angles as above. What is immediately clear is the
departure from the spherical symmetry: both the radial velocity and the density
are not monotonic. Interesting physical insights can be obtained when looking at
the $65^{\circ}$ and $70^{\circ}$ LOS, that are the red and green lines in
Figure~\ref{f7} and Figure~\ref{f8}. These two inclination angles roughly
correspond to the transitional zone from the slow to the fast stream zone.
Observationally, the $65^{\circ}$ red spectral line profile shows much more
blueshifted absorption than the $70^{\circ}$ green profile. However, when
looking at the radial velocity plot in Figure~\ref{f8}, it can be seen that the
absorption at high velocity for $i= 65^{\circ}$ happens close to the origin
($R\sim 150 R_0$, where $R_0=8.8\times 10^{13}$~cm is the inner radius of the accretion disk
and is equal to 3 Schwarzschild radii), where the dynamical instabilities of
the transitional zone of the wind are the highest. At larger radii, for $i=
65^{\circ}$, the radial velocity decreases: the deep absorption signature at
$\upsilon_{out}\sim 0.65\upsilon_{max}$ is due to a portion of the wind (a
'puff', a blob) that will shortly (on dynamical time scale) fall back 
toward the disk plane. On the other
hand, for $i=70^{\circ}$, the maximum of the absorption is reached 
at much lower velocities, and shapes a broad deep trough at $\upsilon_{out}\sim
0.1-0.3\upsilon_{max}$. One can see that most of this absorption happens at
large radii, $R\sim 300 R_0$ , and that this portion of the wind is 
more stable and effectively accelerated to leave the computational domain.

Among the the considered LOS for this particular model, only the largest
inclination angles (70, 75, and 80$^{\circ}$) track portions of the flow that
will successfully escape the system, albeit they show the less blueshifted
absorption profiles (Figure~\ref{f7}). On the contrary, the highly-blueshifted
profile associated to the 65$^{\circ}$ tracks a part of the flow where the gas
is initially accelerated to the highest velocities, but being so close 
to the central continuum source it becomes quickly over-ionized, so loses 
driving force, and falls back to the disk plane. This lack of one-to-one 
correspondency between blueshift and terminal velocity of the wind is again 
contrary to the expectations of a simple radial flow of constant velocity. 
Dynamical effects can be extremely important in shaping the appearance of absorption line profiles,
and should be taken into account when deriving physical properties such as mass
and energy fluxes associated to the wind.

All these considerations can be also visualized in Figure~\ref{f9}, that shows
the map of the ratio between the outflow mass flux computed under the spherical
symmetric assumption and the actual outflow mass flux. The strong blueshifted
absorption seen along the $i = 65^{\circ}$ LOS corresponds to the structure 
close to the origin, where the mass loss rate estimated under spherical 
symmetry is maximally overestimated (up to a factor of 100!). 
For $i = 70^{\circ}$, the discrepancy between the two mass loss rates is much 
smaller, up to an order of magnitude. These simple examples illustrate how 
different (and sometimes counter-intuitively) is a realistic non-spherical, 
unsteady wind from a simple, homogeneous, and steady spherical case.

\section{DISCUSSION}

We have computed synthetic absorption lines against a continuum point source 
for different simulations of accretion disk winds, with the goal of 
investigating the effects of the flow properties on the appearance of 
the line profiles. The used method is as simple and general as possible; scattered 
and emitted radiation is neglected. All the models considered here are Compton-thin in the 
studied LOS, except than for the \citet{2004ApJ...616..688P} line-driven disk wind. 
A proper treatment of Compton-thick LOS requires scattered radiation to be taken into account,
as has been done by means of multidimensional Monte Carlo radiative transfer treatment by  \citet{2010MNRAS.408.1396S}.
Such computations are however extremely time-consuming, hence not easily applicable to the
study of various geometries and LOS for different disk wind models. 

We found that the geometry and kinematic of the flow can imprint their effect 
on the blue edge of the absorption line (Section 3.1). In particular, 
the edge will appear sharp when the radial velocity distribution and 
the opacity distribution along the LOS are $\sim$~constant within the portion 
of the flow that is in resonance with the continuum, i.e. where absorption 
occurs. The edge will instead appear smooth when the radial velocity and 
the opacity have steep profiles along the LOS. 
The extreme sharpness of the blue edge is in part due to the Sobolev approximation
used in this work; in real situations, thermal and turbulent line broadening are expected
to slightly smear the profiles. However, this does not affect the general conclusion of a different level of sharpness of
the blue edge for different radial velocity and opacity distributions across the flow.
We note how UV observations of 
the rare iron low-ionization broad absorption line quasars (FeLoBAL QSOs, that show 
blueshifted absorption in the Fe~\textsc{ii} and Fe~\textsc{iii} transitions) usually reveal 
much sharper blue edges than for the more common high-ionization BAL QSOs 
cases \citep[e.g.,][]{2002ApJS..141..267H}. This might imply a different 
geometry and kinematics of the outflow for these two classes of quasars 
\citep[see for example][]{2000A&A...356L...9L, 2012MNRAS.420.1347F}.

The ionization state of the gas has also important effects in shaping the
absorption line profiles: non-radial solutions to the disk wind problem
offered by hydrodynamical simulations can be steady yet have 
a non-monotonic distribution of the ionization parameter along the flow
and as such can produce complex shapes of the absorption lines
(Section 3.2). In particular, due to the complex distributions of the density
and the radial velocity across the flow,  a higher 
ionization state gas could be slower than a lower-ionization state gas.
Other authors already discussed possible dependences between the velocity
and the ionization state of the wind, in particular referring to the observed X-ray warm
absorber properties. For example, \citet{2011MNRAS.413.1251P} report about a 
positive correlation between the ionization state and the radial 
velocity for the outflowing X-ray absorbing gas observed in NGC 4051, 
thus suggesting agreement with a radial outflow model. On the other hand, \citet{2005ApJ...627..166R} 
suggest the existence of an anti-correlation between $\xi$ and $\upsilon_{out}$ for 
the warm absorber observed against the X-ray continuum of NGC 3783: this observation, along with the asymmetry of 
the absorption line profiles, made the authors conclude that a spherical 
outflow model can not reproduce the observations of NGC 3783, and rather two outflowing 
components modeled using a kinematical radiation-driven wind model are 
required \citep[see also][]{2011RMxAA..47..385R}.
However, for the NGC 3783 case, the authors assume
that the resonance emission lines are narrow, and therefore do
not affect the blue wing of the residual absorption features, while several examples of broad
soft X-ray resonance emission lines have been reported in the literature \citep[e.g.,][]{2002A&A...386..427K, 2007A&A...461..121C,2011MNRAS.413.1251P}. 
The situation is thus far from being clear, and only the launch of future X-ray missions
capable of providing a high spectral resolution over a broad energy range (e.g., the microcalorimeter onboard ASTRO-H) will
allow us to firmly assess the character of such correlations between the ionization state and the velocity of the X-ray absorbing gas.

Not all known disk wind solutions are time independent. Therefore
we also studied synthetic line profiles from highly dynamical models.
We found the dynamics of the wind could strongly contribute to the observed
complexity of the absorption line profiles, especially along the LOS that 
track the most unsteady portions of the flow (Section 3.3). There may not be 
a direct proportionality between the observed velocity shift of the lines 
and the mass outflow rate, and the latter can depend in complex ways on 
the physics and on the geometry of the accretion disk wind.
It appears that detailed simulations of synthetic spectra for accretion disk winds 
are necessary to provide  realistic estimates of the mass outflow rate.
One of the best studied examples is the X-ray absorbing wind observed in PG 1211+143
\citep{2003MNRAS.345..705P}. A mass outflow rate estimate using a spherically
symmetric, constant velocity, geometrically thin shell approximation for the
interpretation of the high-ionization, high-velocity Fe~\textsc{xxv} and Fe~\textsc{xxvi}
absorption lines is $\dot{M}_{out}\sim 8.7M_{\odot}$ yr$^{-1}$ 
\citep{2009MNRAS.397..249P}, that after being corrected for a covering fraction of $\sim 0.4$
translates to $\dot{M}_{out}\sim 3.5M_{\odot}$ yr$^{-1}$. 
By fitting the same data to the models drawn from kinematic disk wind
models and adopting a proper treatment of the radiative transfer in the flow, the mass outflow rate estimate is found to 
be $\dot{M}_{out}\sim 2.1$ M$_{\odot}$ yr$^{-1}$ \citep{2010MNRAS.404.1369S}, i.e. a factor $\sim 40\%$ lower. 
In the future it would be interesting to fit the data of large samples of AGN that display X-ray absorbing outflows
to models derived from hydrodynamical simulations as those illustrated by \citet{2010MNRAS.408.1396S} and \citet{2012arXiv1207.7194S},
to derive reliable mass outflow rates.

Distance estimates for the absorbers often rely  on the assumption of 
a single-zone of  gas photoionized by the central continuum source, 
and exploit the ionization parameter definition $\xi=L/nR^2$ to get 
constraints on the location of the gas based on the observed
density, luminosity, and ionization state. For low-density, low-ionization 
state gas this translates to very large radial distances, of the order of 
the kpc \citep[e.g.,][]{2008ApJ...688..108K,
2009ApJ...706..525M, 2010ApJ...709..611D}. Further assuming a constant 
velocity radial outflow, huge mass outflow rates are derived
(up to several hundreds of solar masses per year). However, as already pointed out by 
\citet{2002ApJ...569..671E}, the derived radial location of the wind is 
strongly lowered in the case of a multi-phase absorber,
where the low-ionization gas is shielded against the ionizing continuum source.
Multiple phases of absorbing gas are exactly what is observed.

It is certainly convenient to think about AGN winds as spherically symmetric,
constant density, geometrically thin shells of gas expanding with uniform
velocity; we then assume photoionization equilibrium to derive constraints on
the distance of the absorbing material from the continuum source. However, the
location of the origin of the wind and of the origin of the absorption lines 
are two different quantities that may or may not coincide. For example, 
a wind can originate from a very small distance from the SMBH, yet 
an absorption line can be formed at larger radii, or over a large radial 
range (e.g. from a few tens up to several hundreds of gravitational radii). 
We have shown that if the wind does not resemble a 'thin shell' geometry but 
is rather radially extended, then the
distribution of the kinematical, ionization, and dynamical properties of the
flow along the LOS all come into play into shaping complex absorption profiles.
Therefore, absorption lines from accretion disk winds form in regions
that only partially cover the source: the covering fraction is
dependent on the geometry, on the velocity, and on the ionization state of 
the gas \citep{2012arXiv1201.3606P}.

 Complex and variable blueshifted absorption profiles are starting to be
 observed in the X-ray spectra of a number of AGN; notably, PG 1126-041 and NGC
 4051 both show variability of complex absorption structures on very short time
 scales in the iron K band \citep{2011A&A...536A..49G,2012MNRAS.423..165P}. X-ray
 broad absorption lines are also observed to be strongly variable in
 shape and strength \citep[][]{2009ApJ...706..644C,2012A&A...544A...2L}. All these
 observational results strongly point against a uniform and steady spherical
 outflow.

\section{CONCLUSIONS\label{conclusions}}
We have considered several physical models of accretion disk winds in AGN,
and computed absorption line profiles predicted by such models. For the simplest
isothermal disk winds, we found
that although the models are non-spherical, they predict shapes of
the synthetic profiles that could have very similar characteristics
for different wind geometries.  Therefore
we conclude that 
the wind geometry is not the main
contributor to the great diversity of the observed line shapes.

More complexities and diversities of the profile shapes can be produced
by facilitating one of the key properties of AGN disk winds: the
non-monotonic distribution of basic wind properties such as velocity,
density, and ionization state. Owing to the complex distribution of these quantities along the LOS, 
multiple and detached absorption troughs can be easily produced in radially extended, non-spherical outflows.
The highly dynamical nature of certain portions of AGN disk winds can also have significant
effects on the mass and energy fluxes estimates, that can be off up to two orders of magnitude with respect
to estimates based on a spherically symmetric, homogeneous, and constant velocity wind.
%

\acknowledgments
We thank Ryuichi Kurosawa, Stuart Sim, Tim Waters, and Fred Hamann for their comments on the manuscript,
Stefan Luketic for providing the streamlines figures, and the referee for useful comments that helped to improve the paper.
We acknowledge support provided by the Chandra awards
TM1-12008X issued by the Chandra X-Ray Observatory Center, 
which is operated by the Smithsonian Astrophysical Observatory for 
and on behalf of NASA under contract NAS 8-39073.
DP also acknowledges the UNLV sabbatical assistance. 

\bibliography{lineprof}

\begin{thebibliography}{73}
\expandafter\ifx\csname natexlab\endcsname\relax\def\natexlab#1{#1}\fi

\bibitem[{{Allen} {et~al.}(2011){Allen}, {Hewett}, {Maddox}, {Richards}, \&
  {Belokurov}}]{2011MNRAS.410..860A}
{Allen}, J.~T., {Hewett}, P.~C., {Maddox}, N., {Richards}, G.~T., \&
  {Belokurov}, V. 2011, \mnras, 410, 860

\bibitem[{{Anderson} {et~al.}(2005){Anderson}, {Li}, {Krasnopolsky}, \&
  {Blandford}}]{2005ApJ...630..945A}
{Anderson}, J.~M., {Li}, Z.-Y., {Krasnopolsky}, R., \& {Blandford}, R.~D. 2005,
  \apj, 630, 945

\bibitem[{{Blandford} \& {Payne}(1982)}]{1982MNRAS.199..883B}
{Blandford}, R.~D., \& {Payne}, D.~G. 1982, \mnras, 199, 883

\bibitem[{{Blustin} {et~al.}(2005){Blustin}, {Page}, {Fuerst},
  {Branduardi-Raymont}, \& {Ashton}}]{2005A&A...431..111B}
{Blustin}, A.~J., {Page}, M.~J., {Fuerst}, S.~V., {Branduardi-Raymont}, G., \&
  {Ashton}, C.~E. 2005, \aap, 431, 111

\bibitem[{{Cappi}(2006)}]{2006AN....327.1012C}
{Cappi}, M. 2006, Astronomische Nachrichten, 327, 1012

\bibitem[{{Cappi} {et~al.}(2009){Cappi}, {Tombesi}, {Bianchi}, {Dadina},
  {Giustini}, {Malaguti}, {Maraschi}, {Palumbo}, {Petrucci}, {Ponti},
  {Vignali}, \& {Yaqoob}}]{2009A&A...504..401C}
{Cappi}, M., {Tombesi}, F., {Bianchi}, S., {et~al.} 2009, \aap, 504, 401

\bibitem[{{Chartas} {et~al.}(2003){Chartas}, {Brandt}, \&
  {Gallagher}}]{2003ApJ...595...85C}
{Chartas}, G., {Brandt}, W.~N., \& {Gallagher}, S.~C. 2003, \apj, 595, 85

\bibitem[{{Chartas} {et~al.}(2002){Chartas}, {Brandt}, {Gallagher}, \&
  {Garmire}}]{2002ApJ...579..169C}
{Chartas}, G., {Brandt}, W.~N., {Gallagher}, S.~C., \& {Garmire}, G.~P. 2002,
  \apj, 579, 169

\bibitem[{{Chartas} {et~al.}(2009){Chartas}, {Saez}, {Brandt}, {Giustini}, \&
  {Garmire}}]{2009ApJ...706..644C}
{Chartas}, G., {Saez}, C., {Brandt}, W.~N., {Giustini}, M., \& {Garmire}, G.~P.
  2009, \apj, 706, 644

\bibitem[{{Costantini} {et~al.}(2007){Costantini}, {Kaastra}, {Arav}, {Kriss},
  {Steenbrugge}, {Gabel}, {Verbunt}, {Behar}, {Gaskell}, {Korista}, {Proga},
  {Quijano}, {Scott}, {Klimek}, \& {Hedrick}}]{2007A&A...461..121C}
{Costantini}, E., {Kaastra}, J.~S., {Arav}, N., {et~al.} 2007, \aap, 461, 121

\bibitem[{{Crenshaw} {et~al.}(2003){Crenshaw}, {Kraemer}, \&
  {George}}]{2003ARA&A..41..117C}
{Crenshaw}, D.~M., {Kraemer}, S.~B., \& {George}, I.~M. 2003, \araa, 41, 117

\bibitem[{{Drew}(1987)}]{1987MNRAS.224..595D}
{Drew}, J.~E. 1987, \mnras, 224, 595

\bibitem[{{Dunn} {et~al.}(2010){Dunn}, {Bautista}, {Arav}, {Moe}, {Korista},
  {Costantini}, {Benn}, {Ellison}, \& {Edmonds}}]{2010ApJ...709..611D}
{Dunn}, J.~P., {Bautista}, M., {Arav}, N., {et~al.} 2010, \apj, 709, 611

\bibitem[{{Everett} {et~al.}(2002){Everett}, {K{\"o}nigl}, \&
  {Arav}}]{2002ApJ...569..671E}
{Everett}, J., {K{\"o}nigl}, A., \& {Arav}, N. 2002, \apj, 569, 671

\bibitem[{{Everett}(2005)}]{2005ApJ...631..689E}
{Everett}, J.~E. 2005, \apj, 631, 689

\bibitem[{{Everett}(2007)}]{2007Ap&SS.311..269E}
---. 2007, \apss, 311, 269

\bibitem[{{Faucher-Gigu{\`e}re} {et~al.}(2012){Faucher-Gigu{\`e}re},
  {Quataert}, \& {Murray}}]{2012MNRAS.420.1347F}
{Faucher-Gigu{\`e}re}, C.-A., {Quataert}, E., \& {Murray}, N. 2012, \mnras,
  420, 1347

\bibitem[{{Foltz} {et~al.}(1987){Foltz}, {Weymann}, {Morris}, \&
  {Turnshek}}]{1987ApJ...317..450F}
{Foltz}, C.~B., {Weymann}, R.~J., {Morris}, S.~L., \& {Turnshek}, D.~A. 1987,
  \apj, 317, 450

\bibitem[{{Font} {et~al.}(2004){Font}, {McCarthy}, {Johnstone}, \&
  {Ballantyne}}]{2004ApJ...607..890F}
{Font}, A.~S., {McCarthy}, I.~G., {Johnstone}, D., \& {Ballantyne}, D.~R. 2004,
  \apj, 607, 890

\bibitem[{{Ganguly} \& {Brotherton}(2008)}]{2008ApJ...672..102G}
{Ganguly}, R., \& {Brotherton}, M.~S. 2008, \apj, 672, 102

\bibitem[{{Gibson} {et~al.}(2009){Gibson}, {Jiang}, {Brandt}, {Hall}, {Shen},
  {Wu}, {Anderson}, {Schneider}, {Vanden Berk}, {Gallagher}, {Fan}, \&
  {York}}]{2009ApJ...692..758G}
{Gibson}, R.~R., {Jiang}, L., {Brandt}, W.~N., {et~al.} 2009, \apj, 692, 758

\bibitem[{{Giustini} {et~al.}(2011){Giustini}, {Cappi}, {Chartas}, {Dadina},
  {Eracleous}, {Ponti}, {Proga}, {Tombesi}, {Vignali}, \&
  {Palumbo}}]{2011A&A...536A..49G}
{Giustini}, M., {Cappi}, M., {Chartas}, G., {et~al.} 2011, \aap, 536, A49

\bibitem[{{Hall} {et~al.}(2002){Hall}, {Anderson}, {Strauss}, {York},
  {Richards}, {Fan}, {Knapp}, {Schneider}, {Vanden Berk}, {Geballe}, {Bauer},
  {Becker}, {Davis}, {Rix}, {Nichol}, {Bahcall}, {Brinkmann}, {Brunner},
  {Connolly}, {Csabai}, {Doi}, {Fukugita}, {Gunn}, {Haiman}, {Harvanek},
  {Heckman}, {Hennessy}, {Inada}, {Ivezi{\'c}}, {Johnston}, {Kleinman},
  {Krolik}, {Krzesinski}, {Kunszt}, {Lamb}, {Long}, {Lupton}, {Miknaitis},
  {Munn}, {Narayanan}, {Neilsen}, {Newman}, {Nitta}, {Okamura}, {Pentericci},
  {Pier}, {Schlegel}, {Snedden}, {Szalay}, {Thakar}, {Tsvetanov}, {White}, \&
  {Zheng}}]{2002ApJS..141..267H}
{Hall}, P.~B., {Anderson}, S.~F., {Strauss}, M.~A., {et~al.} 2002, \apjs, 141,
  267

\bibitem[{{Hamann} {et~al.}(2011){Hamann}, {Kanekar}, {Prochaska}, {Murphy},
  {Ellison}, {Malec}, {Milutinovic}, \& {Ubachs}}]{2011MNRAS.410.1957H}
{Hamann}, F., {Kanekar}, N., {Prochaska}, J.~X., {et~al.} 2011, \mnras, 410,
  1957

\bibitem[{{Hamann} {et~al.}(2008){Hamann}, {Kaplan}, {Rodr{\'{\i}}guez
  Hidalgo}, {Prochaska}, \& {Herbert-Fort}}]{2008MNRAS.391L..39H}
{Hamann}, F., {Kaplan}, K.~F., {Rodr{\'{\i}}guez Hidalgo}, P., {Prochaska},
  J.~X., \& {Herbert-Fort}, S. 2008, \mnras, 391, L39

\bibitem[{{Hamann} {et~al.}(2012){Hamann}, {Simon}, {Rodriguez Hidalgo}, \&
  {Capellupo}}]{2012arXiv1204.3791H}
{Hamann}, F., {Simon}, L., {Rodriguez Hidalgo}, P., \& {Capellupo}, D. 2012,
  ArXiv e-prints

\bibitem[{{Kaastra} {et~al.}(2002){Kaastra}, {Steenbrugge}, {Raassen}, {van der
  Meer}, {Brinkman}, {Liedahl}, {Behar}, \& {de Rosa}}]{2002A&A...386..427K}
{Kaastra}, J.~S., {Steenbrugge}, K.~C., {Raassen}, A.~J.~J., {et~al.} 2002,
  \aap, 386, 427

\bibitem[{{Knigge} {et~al.}(2008){Knigge}, {Scaringi}, {Goad}, \&
  {Cottis}}]{2008MNRAS.386.1426K}
{Knigge}, C., {Scaringi}, S., {Goad}, M.~R., \& {Cottis}, C.~E. 2008, \mnras,
  386, 1426

\bibitem[{{Knigge} {et~al.}(1995){Knigge}, {Woods}, \&
  {Drew}}]{1995MNRAS.273..225K}
{Knigge}, C., {Woods}, J.~A., \& {Drew}, J.~E. 1995, \mnras, 273, 225

\bibitem[{{K{\"o}nigl}(2006)}]{2006MmSAI..77..598K}
{K{\"o}nigl}, A. 2006, \memsai, 77, 598

\bibitem[{{Korista} {et~al.}(2008){Korista}, {Bautista}, {Arav}, {Moe},
  {Costantini}, \& {Benn}}]{2008ApJ...688..108K}
{Korista}, K.~T., {Bautista}, M.~A., {Arav}, N., {et~al.} 2008, \apj, 688, 108

\bibitem[{{Krasnopolsky} {et~al.}(2003){Krasnopolsky}, {Li}, \&
  {Blandford}}]{2003ApJ...595..631K}
{Krasnopolsky}, R., {Li}, Z.-Y., \& {Blandford}, R.~D. 2003, \apj, 595, 631

\bibitem[{{Krongold} {et~al.}(2010){Krongold}, {Binette}, \&
  {Hern{\'a}ndez-Ibarra}}]{2010ApJ...724L.203K}
{Krongold}, Y., {Binette}, L., \& {Hern{\'a}ndez-Ibarra}, F. 2010, \apjl, 724,
  L203

\bibitem[{{Krongold} {et~al.}(2007){Krongold}, {Nicastro}, {Elvis},
  {Brickhouse}, {Binette}, {Mathur}, \&
  {Jim{\'e}nez-Bail{\'o}n}}]{2007ApJ...659.1022K}
{Krongold}, Y., {Nicastro}, F., {Elvis}, M., {et~al.} 2007, \apj, 659, 1022

\bibitem[{{Lamy} \& {Hutsem{\'e}kers}(2000)}]{2000A&A...356L...9L}
{Lamy}, H., \& {Hutsem{\'e}kers}, D. 2000, \aap, 356, L9

\bibitem[{{Lanzuisi} {et~al.}(2012){Lanzuisi}, {Giustini}, {Cappi}, {Dadina},
  {Malaguti}, {Vignali}, \& {Chartas}}]{2012A&A...544A...2L}
{Lanzuisi}, G., {Giustini}, M., {Cappi}, M., {et~al.} 2012, \aap, 544, A2

\bibitem[{{Leighly} {et~al.}(2009){Leighly}, {Hamann}, {Casebeer}, \&
  {Grupe}}]{2009ApJ...701..176L}
{Leighly}, K.~M., {Hamann}, F., {Casebeer}, D.~A., \& {Grupe}, D. 2009, \apj,
  701, 176

\bibitem[{{Long} \& {Knigge}(2002)}]{2002ApJ...579..725L}
{Long}, K.~S., \& {Knigge}, C. 2002, \apj, 579, 725

\bibitem[{{Longinotti} {et~al.}(2010){Longinotti}, {Costantini}, {Petrucci},
  {Boisson}, {Mouchet}, {Santos-Lleo}, {Matt}, {Ponti}, \& {Gon{\c
  c}alves}}]{2010A&A...510A..92L}
{Longinotti}, A.~L., {Costantini}, E., {Petrucci}, P.~O., {et~al.} 2010, \aap,
  510, A92

\bibitem[{{Lovelace} {et~al.}(1991){Lovelace}, {Berk}, \&
  {Contopoulos}}]{1991ApJ...379..696L}
{Lovelace}, R.~V.~E., {Berk}, H.~L., \& {Contopoulos}, J. 1991, \apj, 379, 696

\bibitem[{{Luketic} {et~al.}(2010){Luketic}, {Proga}, {Kallman}, {Raymond}, \&
  {Miller}}]{2010ApJ...719..515L}
{Luketic}, S., {Proga}, D., {Kallman}, T.~R., {Raymond}, J.~C., \& {Miller},
  J.~M. 2010, \apj, 719, 515

\bibitem[{{Ma}(2002)}]{2002MNRAS.335L..99M}
{Ma}, F. 2002, \mnras, 335, L99

\bibitem[{{Matt} {et~al.}(2011){Matt}, {Bianchi}, {Guainazzi}, {Longinotti},
  {Dadina}, {Karas}, {Malaguti}, {Miniutti}, {Petrucci}, {Piconcelli}, \&
  {Ponti}}]{2011A&A...533A...1M}
{Matt}, G., {Bianchi}, S., {Guainazzi}, M., {et~al.} 2011, \aap, 533, A1

\bibitem[{{McKernan} {et~al.}(2007){McKernan}, {Yaqoob}, \&
  {Reynolds}}]{2007MNRAS.379.1359M}
{McKernan}, B., {Yaqoob}, T., \& {Reynolds}, C.~S. 2007, \mnras, 379, 1359

\bibitem[{{Miniutti} {et~al.}(2007){Miniutti}, {Ponti}, {Dadina}, {Cappi}, \&
  {Malaguti}}]{2007MNRAS.375..227M}
{Miniutti}, G., {Ponti}, G., {Dadina}, M., {Cappi}, M., \& {Malaguti}, G. 2007,
  \mnras, 375, 227

\bibitem[{{Misawa} {et~al.}(2007){Misawa}, {Eracleous}, {Charlton}, \&
  {Kashikawa}}]{2007ApJ...660..152M}
{Misawa}, T., {Eracleous}, M., {Charlton}, J.~C., \& {Kashikawa}, N. 2007,
  \apj, 660, 152

\bibitem[{{Moe} {et~al.}(2009){Moe}, {Arav}, {Bautista}, \&
  {Korista}}]{2009ApJ...706..525M}
{Moe}, M., {Arav}, N., {Bautista}, M.~A., \& {Korista}, K.~T. 2009, \apj, 706,
  525

\bibitem[{{Pounds} \& {Reeves}(2009)}]{2009MNRAS.397..249P}
{Pounds}, K.~A., \& {Reeves}, J.~N. 2009, \mnras, 397, 249

\bibitem[{{Pounds} {et~al.}(2003){Pounds}, {Reeves}, {King}, {Page}, {O'Brien},
  \& {Turner}}]{2003MNRAS.345..705P}
{Pounds}, K.~A., {Reeves}, J.~N., {King}, A.~R., {et~al.} 2003, \mnras, 345,
  705

\bibitem[{{Pounds} \& {Vaughan}(2011)}]{2011MNRAS.413.1251P}
{Pounds}, K.~A., \& {Vaughan}, S. 2011, \mnras, 413, 1251

\bibitem[{{Pounds} \& {Vaughan}(2012)}]{2012MNRAS.423..165P}
---. 2012, \mnras, 423, 165

\bibitem[{{Proga}(2005)}]{2005ApJ...630L...9P}
{Proga}, D. 2005, \apjl, 630, L9

\bibitem[{{Proga}(2007)}]{2007ASPC..373..267P}
{Proga}, D. 2007, in Astronomical Society of the Pacific Conference Series,
  Vol. 373, The Central Engine of Active Galactic Nuclei, ed. {L.~C.~Ho \&
  J.-W.~Wang}, 267

\bibitem[{{Proga} \& {Kallman}(2004)}]{2004ApJ...616..688P}
{Proga}, D., \& {Kallman}, T.~R. 2004, \apj, 616, 688

\bibitem[{{Proga} {et~al.}(2012){Proga}, {Rodr{\'{\i}}guez Hidalgo}, \&
  {Hamann}}]{2012arXiv1201.3606P}
{Proga}, D., {Rodr{\'{\i}}guez Hidalgo}, P., \& {Hamann}, F. 2012, ArXiv
  e-prints

\bibitem[{{Proga} {et~al.}(2000){Proga}, {Stone}, \&
  {Kallman}}]{2000ApJ...543..686P}
{Proga}, D., {Stone}, J.~M., \& {Kallman}, T.~R. 2000, \apj, 543, 686

\bibitem[{{Ram{\'{\i}}rez}(2011)}]{2011RMxAA..47..385R}
{Ram{\'{\i}}rez}, J.~M. 2011, 47, 385

\bibitem[{{Ram{\'{\i}}rez} {et~al.}(2005){Ram{\'{\i}}rez}, {Bautista}, \&
  {Kallman}}]{2005ApJ...627..166R}
{Ram{\'{\i}}rez}, J.~M., {Bautista}, M., \& {Kallman}, T. 2005, \apj, 627, 166

\bibitem[{{Reeves} {et~al.}(2003){Reeves}, {O'Brien}, \&
  {Ward}}]{2003ApJ...593L..65R}
{Reeves}, J.~N., {O'Brien}, P.~T., \& {Ward}, M.~J. 2003, \apjl, 593, L65

\bibitem[{{Risaliti} {et~al.}(2005){Risaliti}, {Bianchi}, {Matt}, {Baldi},
  {Elvis}, {Fabbiano}, \& {Zezas}}]{2005ApJ...630L.129R}
{Risaliti}, G., {Bianchi}, S., {Matt}, G., {et~al.} 2005, \apjl, 630, L129

\bibitem[{{Rodr{\'{\i}}guez Hidalgo} {et~al.}(2011){Rodr{\'{\i}}guez Hidalgo},
  {Hamann}, \& {Hall}}]{2011MNRAS.411..247R}
{Rodr{\'{\i}}guez Hidalgo}, P., {Hamann}, F., \& {Hall}, P. 2011, \mnras, 411,
  247

\bibitem[{{Shlosman} \& {Vitello}(1993)}]{1993ApJ...409..372S}
{Shlosman}, I., \& {Vitello}, P. 1993, \apj, 409, 372

\bibitem[{{Sim} {et~al.}(2010{\natexlab{a}}){Sim}, {Miller}, {Long}, {Turner},
  \& {Reeves}}]{2010MNRAS.404.1369S}
{Sim}, S.~A., {Miller}, L., {Long}, K.~S., {Turner}, T.~J., \& {Reeves}, J.~N.
  2010{\natexlab{a}}, \mnras, 404, 1369

\bibitem[{{Sim} {et~al.}(2012){Sim}, {Proga}, {Kurosawa}, {Long}, {Miller}, \&
  {Turner}}]{2012arXiv1207.7194S}
{Sim}, S.~A., {Proga}, D., {Kurosawa}, R., {et~al.} 2012, ArXiv e-prints

\bibitem[{{Sim} {et~al.}(2010{\natexlab{b}}){Sim}, {Proga}, {Miller}, {Long},
  \& {Turner}}]{2010MNRAS.408.1396S}
{Sim}, S.~A., {Proga}, D., {Miller}, L., {Long}, K.~S., \& {Turner}, T.~J.
  2010{\natexlab{b}}, \mnras, 408, 1396

\bibitem[{{Tombesi} {et~al.}(2010){Tombesi}, {Cappi}, {Reeves}, {Palumbo},
  {Yaqoob}, {Braito}, \& {Dadina}}]{2010A&A...521A..57T}
{Tombesi}, F., {Cappi}, M., {Reeves}, J.~N., {et~al.} 2010, \aap, 521, A57

\bibitem[{{Turner} {et~al.}(2008){Turner}, {Reeves}, {Kraemer}, \&
  {Miller}}]{2008A&A...483..161T}
{Turner}, T.~J., {Reeves}, J.~N., {Kraemer}, S.~B., \& {Miller}, L. 2008, \aap,
  483, 161

\bibitem[{{Turnshek}(1984)}]{1984ApJ...280...51T}
{Turnshek}, D.~A. 1984, \apj, 280, 51

\bibitem[{{Turnshek} {et~al.}(1988){Turnshek}, {Grillmair}, {Foltz}, \&
  {Weymann}}]{1988ApJ...325..651T}
{Turnshek}, D.~A., {Grillmair}, C.~J., {Foltz}, C.~B., \& {Weymann}, R.~J.
  1988, \apj, 325, 651

\bibitem[{{Vivek} {et~al.}(2012){Vivek}, {Srianand}, {Mahabal}, \&
  {Kuriakose}}]{2012MNRAS.421L.107V}
{Vivek}, M., {Srianand}, R., {Mahabal}, A., \& {Kuriakose}, V.~C. 2012, \mnras,
  421, L107

\bibitem[{{Waters} \& {Proga}(2012)}]{2012arXiv1207.7348W}
{Waters}, T.~R., \& {Proga}, D. 2012, ArXiv e-prints

\bibitem[{{Woods} {et~al.}(1996){Woods}, {Klein}, {Castor}, {McKee}, \&
  {Bell}}]{1996ApJ...461..767W}
{Woods}, D.~T., {Klein}, R.~I., {Castor}, J.~I., {McKee}, C.~F., \& {Bell},
  J.~B. 1996, \apj, 461, 767

\bibitem[{{Wu} {et~al.}(2010){Wu}, {Charlton}, {Misawa}, {Eracleous}, \&
  {Ganguly}}]{2010ApJ...722..997W}
{Wu}, J., {Charlton}, J.~C., {Misawa}, T., {Eracleous}, M., \& {Ganguly}, R.
  2010, \apj, 722, 997

\end{thebibliography}
\end{document}